\documentclass[a4paper,11pt]{article}
\usepackage{pos}

\def\beq{\begin{equation}}
\def\eeq{\end{equation}}
\def\beqn{\begin{eqnarray}}
\def\eeqn{\end{eqnarray}}

\newcommand{\cc}[2]{c{#1\atopwithdelims[]#2}}

\def\NPB#1#2#3{{\it Nucl.\ Phys.}\/ {\bf B#1} (#2) #3}
\def\PLB#1#2#3{{\it Phys.\ Lett.}\/ {\bf B#1} (#2) #3}

\def\PRD#1#2#3{{\it Phys.\ Rev.}\/ {\bf D#1}  (#2) #3}
\def\PRL#1#2#3{{\it Phys.\ Rev.\ Lett.}\/ {\bf #1} (#2) #3}

\def\EJP#1#2#3{ {\it Eur.\ Phys.\ Jour.}\/ {\bf C#1} (#2) #3}
\def\JHEP#1#2#3{ {\it JHEP}\/ {\bf #1} (#2) #3}

\def\etal{{\it et al\/}}

\def\AEF{A.E. Faraggi}

\title{Mirror Symmetry and Spinor--Vector Duality:\\
\medskip
A Top--Down Approach to the Swampland Program}

\author*[a]{Alon E. Faraggi}

\affiliation[a]{Department of Mathematical Sciences, university of Liverpool, 
  Liverpool L69 7ZL, United Kingdom}

\emailAdd{alon.faraggi@liv.ac.uk}

\abstract{Mirror symmetry is one of the celebrated developments in pure mathematics that arose from an initial observation in worldsheet string 
constructions. The profound implications of mirror symmetry in the 
Effective Field Theory (EFT) limit of string compactifications was 
subsequently understood. In particular, it proved to be an exceptionally 
useful tool in the field of enumerative geometry. Spinor--Vector Duality (SVD) is an extension of mirror symmetry that can be readily understood in terms of the moduli parameters of 
toroidal heterotic--string compactifications, which include the metric, the anti--symmetric 
ternsor field and the Wilson--line moduli. While mirror symmetry corresponds to 
maps of the internal moduli parameters, {\i.e.} the metric and the anti--symmetric tensor 
field, SVD corresponds to maps of the Wilson--line moduli. Similar to mirror symmetry the 
imprint of SVD in the EFT limit can serve as a tool to study the properties of complex 
manifolds with vector--bundles. Spinor--Vector Duality motivates a top--down approach
to the "Swampland" program, by studying the imprint of the symmetries of the 
worldsheet ultra--violet complete string constructions in the EFT limit. It is conjectured
that SVD provides a demarcation line between (2,0) EFTs that possess an ultra--violet
complete embedding in string theory versus those that do not. 
}

\FullConference{Proceedings of the Corfu Summer Institute 2024 "School and Workshops on Elementary Particle Physics and Gravity" (CORFU2024)\
12 - 26 May, and 25 August - 27 September, 2024\\
Corfu, Greece\\}

\begin{document}
\maketitle

\section{Introduction}

String theory provides a self--consistent perturbative framework for the synthesis of
quantum mechanics and gravity. Its self--consistency conditions imply the
existence of the gauge and matter sectors that arise in the Standard Model of particle
physics. As such, it provides a well--defined perturbative particle theory with well--defined
physics attributes. In this respect, string theory is a mundane
extension of the concept of a point particle theory. In the latter,
we parametrise the paths of particles with one worldline parameter,
whereas in the former, we parametrise the path of particles with
two worldsheet parameters. There is nothing sacred about the single
parameter paramtrisation of the paths of elementary particles. What
we learn from string theory is that a consistent theory of quantum gravity
requires a departure from the worldline parametrisation which is
adopted in conventional quantum field theories.

It is important to note that the Grand Unified Theory paradigm,
and ultimately the unification of the Standard Quantum Field Theory Model
with Gravity, is favoured by the experimental data.
The main merit of string theory is that it gives rise to the basic
ingredients in the Standard Model of particle physics and provides
the framework for the development of a phenomenological approach to
quantum gravity. Indeed, since the late 1980s the fermionic 
$Z_2\times Z_2$ orbifold provide benchmark models for detailed 
studies of the phenomenology of the Standard Model and Unification.
Among these are: 
\begin{itemize}
\item Minimal Standard Heterotic--String Mode \cite{fny}.
\item Top quark mass at $\sim 175-180$GeV \cite{top}. 
\item Fermion masses and CKM mixing \cite{fmmckm}. 
\item Neutrino masses \cite{nmasses}.
\item Gauge coupling unification \cite{gcu}.
\item Supersymmetry breaking \cite{FKP}. 
\item Proton stability \cite{ps}.
\item Moduli fixing \cite{moduli}.
\item Classification ... 
\cite{SO10classi, svddiscovery, psclassi, fsuclassi, SMclassi, LRSclassi, fmp}. 
\end{itemize}
The fermionic $Z_2\times Z_2$ orbifold gives rise to a large space of 
three generation models with some unbroken $SO(10)$ GUT subgroup, and viable 
Higgs states to reduce the symmetry to that of the Standard Model.
while the early quasi--realistic free fermion models corresponded to isolated
examples \cite{fsu5, fny, slm, alr}, over the past 20+ years systematic classication
methods were developed that enables the scan of billions of string vacua
\cite{SO10classi, svddiscovery, psclassi, fsuclassi, SMclassi, LRSclassi, fmp}
and the analysis of their spectra and properties. One of the clear
achievements of the classification methodology is the discovery of
Spinor--Vector Duality \cite{svddiscovery}. 
Spinor--Vector Dulaity is an extension of 
the celebrated Mirror Symmetry \cite{GreenePlesser, CLS, COGP} in string theory and Calabi--Yau (CY) 
compactifications. The main interest in Mirror Symmetry is from a
purely mathematical point of view as it provides a powerful tool to analyse 
the properties of the complex manifolds. Similarly, SVD provides a
tool that can be used to analyse the properties of CY manifolds 
with vector bundles. 

\section{Free Fermionic Classification Method}

Spinor--Vector Duality was first observed in the free fermionic formulation (FFF)
\cite{fff} by using the systematic classification methods of the
heterotic--string in four dimensions. 
In the free fermionic formulation all the
degrees of freedom required to cancel the conformal anomaly are
worldsheet Majorana--Weyl free fermions. 
These fermions are free only at a specific point in the moduli
space. The moduli fields that enable deformation from the 
free fermion point are present in the spectrum.
The constructions are mathematically
equivalent to bosonic compactifications on six dimensional tori,
with worldsheet Thirring interactions, which are equivalent to
the exact marginal deformations in the bosonic models.
In four
dimensional models in the light--cone gauge the total number of
worldsheet fermions is twenty left--moving and forty--four
right--moving real two dimensional fermions. Eight of the 
left--moving fermions correspond to the Ramond--Neveu--Schwarz
fermions of the ten dimensional superstring.
The additional twelve worldsheet real fermions correspond to 
the six left--moving compactified coordinates, and 
twelve right--moving real fermions correspond to 
the six compactified dimensions
on the bosonic side. The additional thirty--two real
fermions are combined into sixteen complex fermions that
give rise to the rank sixteen gauge symmetry of the 
ten dimensional heterotic--string. The sixty--four 
real worldsheet fermions are denoted by:
\leftline{~~~${\underline{{\hbox{Left-Movers}}}}$:~~~~~~~~~~~~~~~~~~~~~~~~
~~~~$\psi^\mu,~~{ \chi_i},~~{ y_i,~~\omega_i}~~~~(\mu=1,2,~i=1,\cdots,6)$}
\vspace{4mm}
\leftline{~~~${\underline{{\hbox{Right-Movers}}}}$}
$${\bar\phi}_{A=1,\cdots,44}=
\begin{cases}
~~{ {\bar y}_i~,~ {\bar\omega}_i} & i=1,{\cdots},6\cr
  & \cr
~~{ {\bar\eta}_i} & i=1,2,3~~\cr
~~{ {\bar\psi}_{1,\cdots,5}} & \cr
~~{{\bar\phi}_{1,\cdots,8}}  & 
\end{cases}
$$
\noindent
where ${\bar\psi}^{1,\cdots,5}$ give rise the $SO(10)$ GUT symmetry
and ${\bar\phi}^{1,\cdots,8}$ produce the
hidden sector gauge symmetry; ${\bar\eta}^{1,2,3}$ 
give rise to three $U(1)$ observable symmetries;
$\{y,\omega\vert{\bar y},{\bar\omega}\}^{1,\cdots,6}$
correspond to the internal six compactified dimensions; 
String models in the free fermionic construction are defined
in terms of a set of boundary condition basis vectors, 
which denote the phases of the fermions around the noncontractible
loops of the vacuum to vacuum amplitude, and the Generalised GSO (GGSO)
projection coefficients of the one--loop partition function \cite{fff}.

The free fermionic model building tools were used to develop
systematic classification of $Z_2\times Z_2$ heterotic--string
orbifolds. The method was initially developed
to classify vacua with unbroken $SO(10)$ symmetry
and with respect to the number of spinorial $16$, anti--spinorial 
$\overline{16}$, and vectorial $10$, representaions of $SO(10)$
\cite{SO10classi}. 
This led to the discovery of Spinor--Vector Duality (SVD) in the space of
fermionic $Z_2\times Z_2$ orbifold compactification, where the
duality transformation is induced by exchange of GGSO phases.
In the free fermion classification method with unbroken $SO(10)$
gauge symmetry, the string models are
produced with a fixed set of basis vectors, 
consisting of twelve basis vectors, 
$
B=\{v_1,v_2,\dots,v_{12}\}, 
$
\begin{eqnarray}
v_1={\bf1}&=&\{\psi^\mu,\
\chi^{1,\dots,6},y^{1,\dots,6}, \omega^{1,\dots,6}~~~|
~~~\bar{y}^{1,\dots,6},\bar{\omega}^{1,\dots,6},
\bar{\eta}^{1,2,3},
\bar{\psi}^{1,\dots,5},\bar{\phi}^{1,\dots,8}\},\nonumber\\
v_2=S&=&\{\psi^\mu,\chi^{1,\dots,6}\},\nonumber\\
v_{3}=z_1&=&\{\bar{\phi}^{1,\dots,4}\},\nonumber\\
v_{4}=z_2&=&\{\bar{\phi}^{5,\dots,8}\},
\label{basis}\\
v_{4+i}=e_i&=&\{y^{i},\omega^{i}|\bar{y}^i,\bar{\omega}^i\}, \ i=1,\dots,6,
~~~~~~~~~~~~~~~~~~~~~N=4~~{\rm Vacua}
\nonumber\\
& & \nonumber\\
v_{11}=b_1&=&\{\chi^{34},\chi^{56},y^{34},y^{56}|\bar{y}^{34},
\bar{y}^{56},\bar{\eta}^1,\bar{\psi}^{1,\dots,5}\},
~~~~~~~~N=4\rightarrow N=2\nonumber\\
v_{12}=b_2&=&\{\chi^{12},\chi^{56},y^{12},y^{56}|\bar{y}^{12},
\bar{y}^{56},\bar{\eta}^2,\bar{\psi}^{1,\dots,5}\},
~~~~~~~~N=2\rightarrow N=1. \nonumber
\end{eqnarray}
The first ten basis vectors preserve $N=4$ spacetime supersymmetry.
The $e_i$ basis vectors correspond to order 2 shifts in the internal 
dimensions. The $z_{1,2}$ reduce the untwisted hidden sector gauge group to
$SO(8)\times SO(8)$. 
The vectors $b_1$ and $b_2$ are the $Z_2\times Z_2$ orbifold twists. 
The third twisted sector is obtained as the combination
$b_3= b_1+b_2+x$, where the $x$--sector is given by
\beq
x= {\bf1} +S + \sum_{i=1}^6 e_i +\sum_{k=1}^2 z_k =
\{{\bar\psi}^{1,\cdots, 5}, {\bar\eta}^{1,2,3}\}.
\label{xmap}
\eeq
The space of string models is spanned by 
varying the independent GGSO projection coefficients, 
$$%
\bordermatrix{%
         &1   &  S  &  e_1  &   e_2   &  e_3   &  e_4  &   e_5  &   e_6%
&  z_1   &  z_2   &  b_1  &   b_2 \cr%
   1   & -1   &  -1 &  \pm  &  \pm  &  \pm  &  \pm  &  \pm   &  \pm  & \pm  &%
 \pm   & \pm  & \pm \cr%
   S~   &   &    & -1 & -1 & -1 & -1 & -1 & -1 & -1 & -1 &  1 & 1 \cr%
  e_1~  &   &    &    &\pm &\pm &\pm &\pm &\pm &\pm &\pm &\pm &\pm\cr%
  e_2~  &   &    &    &    &\pm &\pm &\pm &\pm &\pm &\pm &\pm &\pm\cr%
  e_3~  &   &    &    &    &    &\pm &\pm &\pm &\pm &\pm &\pm &\pm\cr%
  e_4~  &   &    &    &    &    &    &\pm &\pm &\pm &\pm &\pm &\pm\cr%
  e_5~  &   &    &    &    &    &    &    &\pm &\pm &\pm &\pm &\pm\cr%
  e_6~  &   &    &    &    &    &    &    &    &\pm &\pm &\pm &\pm\cr%
  z_1~  &   &    &    &    &    &    &    &    &    &\pm &\pm &\pm\cr%
  z_2~  &   &    &    &    &    &    &    &    &    &    &\pm &\pm\cr%
  b_1~  &   &    &    &    &    &    &    &    &    &    &    &\pm\cr%
  b_2~  &   &    &    &    &    &    &    &    &    &    &    &   \cr%
  }~. %
$$%
The diagonal phases and below are fixed by modular invariance.
Additional fixed phases are set by the overall chirality and 
by requiring $N=1$ spacetime supersymmetry.
A specific choice of the 55, $\pm1$ phases
corresponds to a distinct string vacuum.
Distinct selection phases are generated by a random generator
and the spectrum is analysed for each choice. In ref. \cite{smt}
it was shown that utilisation of the Satisfiabilty Modulo Theories
algorithm can reduce the computational running time by three 
orders of magnitude in some casee. 

\section{Mirror Symmetry}

To see the mirror symmetry operation in the case of fermionic $Z_2\times Z_2$
orbifold, it is instrumental to enhance the untwisted $SO(10)\times U(1)$ 
symmetry to $E_6$.
This is obtained by retaining the spacetime vector bosons 
from the $x$--sector,
in the $16\oplus\overline{16}$ of $SO(10)$, which completes the adjoint 
representation $45\oplus1$ to the $78$ adjoint representation of $E_6$.
The chiral matter states in the $27$ and $\overline{27}$ of $E_6$
are obtained from sectors $b_j\oplus b_j+X, ~j=1,2,3$. 
The mirror symmetry, which exchanges the Euler characteristic,
$$
\frac{\chi}{2} = \# (27-\overline{27})~~\longrightarrow~~-\frac{\chi}{2}, 
$$
as well as the Complex and K\"ahler structure moduli, arises
from the exchange,
\beq
\cc{b_1}{b_2} = +1 \longrightarrow \cc{b_1}{b_2}= -1~.
\label{mirrorexchange}
\eeq
This exchange in the free fermion construction correspond to an 
exchange of a discrete torsion in the partition function. 
In \cite{vwmirror} Vafa and Witten discussed mirror symmetry in 
$Z_2\times Z_2$ orbifolds in terms of exchange of a discrete
torsion. 

\section{Spinor--Vector Duality}

\begin{figure}[h]
\includegraphics[width=14pc]{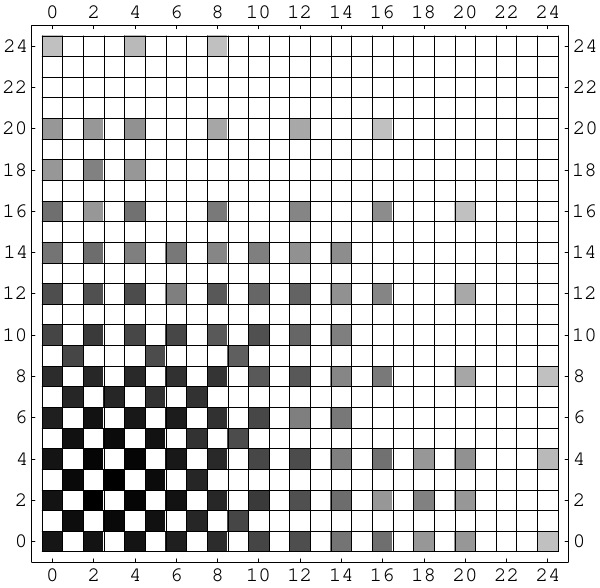}\hspace{0.5pc}%
\begin{minipage}[b]{20pc}
\caption{
\label{den}
Density plot showing the spinor--vector duality in the space of fermionic
$Z_2\times Z_2$ heterotic--string orbifolds. The figure displays the number 
of models with a number of $({16}+\overline{16})$ and a number
{10} representations of $SO(10)$. It is symmetric under exchange of 
rows and columns, reflecting the Spinor--Vector Duality that 
underlies the entire space of models.}
\end{minipage}
\end{figure}

The computerised free fermionic classification method enabled
the discovery of Spinor--Vector Duality (SVD)
that underlies the space of (2,0) heterotic--string
compactifications. SVD is akin to mirror symmetry.
In terms of the toroidal orbifold moduli, which include the
metric, the antisymmetric tensor field, and the Wilson line moduli, 
mirror symmetry corresponds to mappings of the internal moduli, {\it i.e.}
the metric and the antisymmetric tensor field, whereas SVD arises from
exchanges of the Wilson line moduli. 
The case with unbroken $SO(10)$ gauge symmetry allows for a 
complete generation of all the vacua, which facilitated the 
observation of the underlying Spinor--Vector Duality.
SVD operates in vacua in which the $N=4$ spacetime supersymmetry is
reduced to $N=2$, or $N=1$, by a $Z_2$ twist, or $Z_2\times Z_2$ 
twists, of the internal compactified coordinates, respectively. 
The gauge group arising from the ten dimensional $E_8\times E_8$ 
symmetry depends on the action of a Wilson line. In the absence
of a Wilson line the gauge symmetries are $E_8\times E_8$ in the $N=4$
case, $E_7\times SU(2)\times E_8$ in the $N=2$ case, and
$E_6\times U(1)^2\times E_8$ in the $N=1$ case.
The $Z_2$  twists of the internal coordinates generate
twisted sectors that give rise to massless states in
the $56$ representation of $E_7$, and $27$ and $\overline{27}$
representation of $E_6$. 
The Wilson line breaks the gauge groups in these cases
to $SO(16)\times SO(16)$, $SO(12)\times SU(2)\times SU(2)\times SO(16)$
and $SO(10)\times U(1)^3\times SO(16)$, respectively.
The $56$, and $27$ and $\overline{27}$, representations of 
the $E_7$ and $E_6$ gauge groups are
decomposed into spinorial and vectorial representations
of the $SO(12)\times SU(2)$ and $SO(10)\times SO(2)$ subgroups, 
which are $(32,1)$, $(32^\prime,1)$ and $(12, 2)$ 
in the first case and
$16$, $\overline{16}$ and $10$ in the second. 
The Spinor--Vector Duality (SVD) exchanges
the spinor and vector representations.
In the $N=1$ case, the $27$ and $\overline{27}$
representations of $E_6$ decompose as
\beq
27=16_{+1/2}+10_{-1}+1_{+2}~~~~~
{\rm and}~~~~~ \overline{27}=\overline{16}_{-1/2}+10_{+1}+1_{-2}
\label{e627tso10}
\eeq
under $SO(10)\times U(1)$. It is observed that in
$E_6$ for every $16$ multiplet and $\overline{16}$ multiplet, 
there is a vectorial $10$ multiplet.
Hence, string vacua with $E_6$ symmetry are self--dual
under the exchange of the total number of $16\oplus\overline{16}$
multiplets with the total number of
vectorial $10$ mutltiplets. 
There is a remnant of this symmetry when the $E_6$ gauge group
is broken to $SO(10)\times U(1)$. 
For a string vacuum with a number of spinorial and anti--spinorial $SO(10)$ representations, $\#_1(16+\overline{16})$
and a $\#_2(10)$ of vectorial representations, there is a
dual vacuum in which $\#_1\longleftrightarrow\#_2$.
The duality was first observed by counting the total number 
of vacua with a number of $(16+\overline{16})$ representations
versus the total number of vacua with a number of $10$
representaions, in one of the twisted planes of the
$Z_2\times Z_2$ orbifold, which is shown in the table below.
\begin{center}
\begin{tabular}{|ccc|ccc|ccc|c|}
\hline
      & First Plane & &    & Second plane & & & Third Plane & & \\
\hline
$s$ &${\bar s}$& $v$ &$s$&${\bar s}$&$v$&$s$&${\bar s}$&$v$& \# of models \\
\hline
2 & 0 & 0 &    0 & 0 & 0 &    0 & 0 & 0 & 1325963712 \\
0 & 2 & 0 &    0 & 0 & 0 &    0 & 0 & 0 & 1340075584 \\
1 & 1 & 0 &    0 & 0 & 0 &    0 & 0 & 0 & 3718991872 \\
\hline
0 & 0 & 2 &    0 & 0 & 0 &    0 & 0 & 0 & 6385031168 \\
\hline
\hline
\end{tabular}
\end{center}
Figure \ref{den} displays the Spinor--Vector Duality in the full space
of fermionic $Z_2\times Z_2$ orbifolds. It is noted that the figure
is symmetric under the exchange of rows and columns, reflecting the 
underlying Spinor--Vector Duality. The line along the diagonal are 
self--dual models. However, here the scan is restricted to vacua 
on which the $SO(10)\times U(1)$ symmetry is not enhanced to $E_6$.
Hence, the self--dual vacua are those in which the chiral spectrum
comes in complete $E_6$ multiplets but the gauge symmetry is
$SO(10)\times U(1)$, with the $U(1)$ being anomaly free. This was exploited
in \cite{frzprime} to construct a string model
that allows for an extra $U(1)$ symmetry to remain unbroken 
down to low scales. 

The Spinor--Vector Duality is similar to $T$--duality.
The $E_6$ enhanced symmetry point is
self--dual under the Spinor--Vector Duality.
The SVD can also be understood
in terms of a spectral flow operator 
on the bosonic side of the heterotic--string \cite{svddiscovery, FFMT}. 
Vacua with $E_6$ gauge symmetry have $(2,2)$ worldsheet
supersymmetry. In the free fermionic models with basis set eq.
(\ref{basis}), the $S$--vectors is a spectral flow operator 
on the supersymmetric side that exchanges between
sectors that produce spacetime bosons and fermions. Likewise,
on the bosonic side, in the vacua with $(2,2)$ worldsheet supersymmetry
and enhanced $E_6$ gauge symmetry there is a spectral flow
operator that exchanges between the sectors that produce
spinorial and vectorial components of $SO(10)\times U(1)$
in the chiral representaions of $E_6$
\cite{svddiscovery, FFMT}. In vacua with enhanced $E_6$ symmetry,
the spectral flow operator on the bosonic side
mixes between the spinorial and vectorial representations
in the breaking of $E_6$ under $SO(10)\times U(1)$.
In vacua in which $E_6~\rightarrow SO(10)\times U(1)$,
the $N=2$ worldsheet supersymmetry on the bosonic side is
broken, and the spectral flow operator induces the map between the 
spinor--vector dual vacua \cite{svddiscovery,FFMT}.

\begin{figure}[!]
	\centering
	\includegraphics[width=100mm]{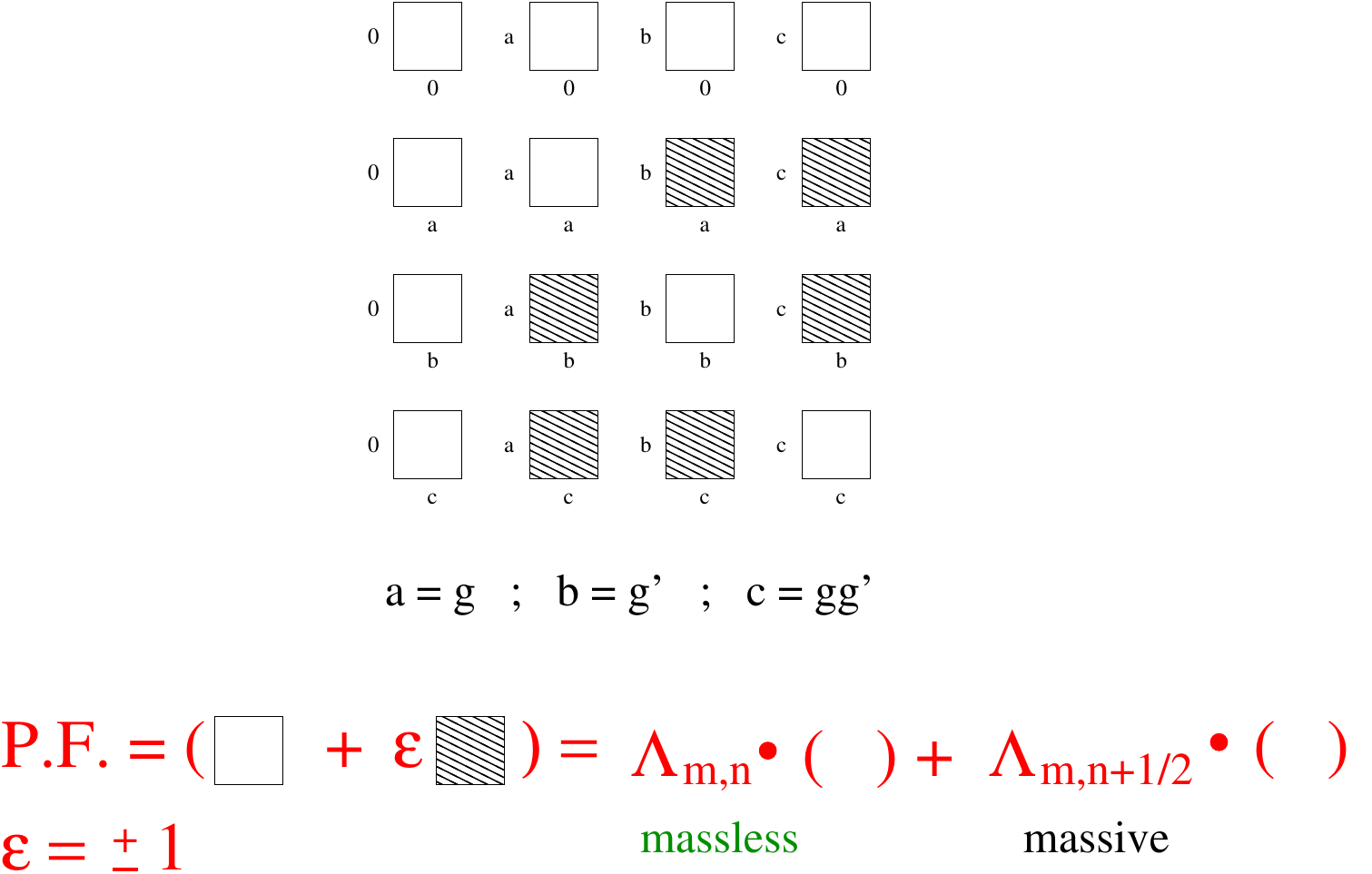}
	\caption{{
            The
            $Z_2\times Z_2^\prime$ partition function of the $g^\prime$--twist and $g$ Wilson line
            with discrete torsion $\epsilon=\pm1$. 
}
}
	\label{z2z2svd}
\end{figure}

Similar to the role of the discrete torsion in the case of 
mirror symmetry in $Z_2\times Z_2$ orbifolds \cite{vwmirror}, 
the Spinor--Vector duality can be realised in terms of a discrete
torsion \cite{CJFKR, aft, FFMT}.
This was reviewed in some detail in \cite{universe},
and is only briefly summarised here.
To realise the SVD in terms of discrete torsion,
the representation of the string vacuum is
translated from the free fermionic to a free bosonic
representation in which the internal
dimensions are compactified on a flat 
six dimensional torus \cite{afgm}, producing the $Z_+$
partition function of the $E_8\times E_8$ heterotic--string
compactified to four dimensions on a flat torus. 
To this model are added two $Z_2$ actions, 
The first $Z_2$ is a twist of the internal coordinates given by 
\beq
{g^\prime}:(x_{4},x_{5},x_{6},x_7,x_8,x_9)
\longrightarrow
(-x_{4},-x_{5},-x_{6},-x_7,+x_8,+x_9). 
\label{z2twist}
\eeq
The second $Z_2$ action can be interpreted as a Wilson line
in along a circle in the $X_9$ direction \cite{FFMT},
$$
g~: (0^7,1|1, 0^7)  ~\rightarrow~
E_8\times E_8\rightarrow SO(16)\times SO(16).\label{wilsonline}
$$
The $Z_2$ twist in the internal space breaks $N=4\rightarrow N=2$ spacetime
supersymmetry and $E_8\rightarrow E_7\times SU(2)$, or
with the inclusion of the Wilson line 
$SO(16)\rightarrow SO(12)\times SO(4)$.
The orbifold partition function is
$${Z~=~
\left({Z_+\over{Z_g\times Z_{g^{\prime}}}}\right)~=~
\left[{{(1+g)}\over2}{{(1+g^\prime)}\over2}\right]~Z_+}.$$
The partition function contains an untwisted sector and three twisted sectors.
Its schematic form is shown in figure \ref{z2z2svd}.
The sign of the discrete torsion $\epsilon=\pm$ fixes
that either the vectorial states, 
or the spinorial states, are massless.
The spinor--vector duality arises due to the exchange of the
discrete torsion
$$\epsilon=+1\rightarrow \epsilon =-1$$
in the $Z_2\times Z_2^\prime$ partition function.
This is similar to mirror symmetry in the $Z_2\times Z_2$
orbifold of ref. \cite{vwmirror}, where the mirror symmetry
arises due to
the exchange of the discrete torsion between the two 
$Z_2$ orbifold twists.
This example reveals the inner workings of the SVD transformation. 
As shown in table above and fig. \ref{den},
the Spinor--Vector duality works in the 
larger space of string vacua with $N=2$ and $N=1$ spacetime supersymmetry
\cite{svddiscovery}.

The free fermionic analysis obscures the role of 
the Wilson--line moduli fields, but proves
in terms of the GGSO projection coefficients of the one--loop partition function
that the SVD always exists in this space of vacua
\cite{svddiscovery}.
The bosonic analysis \cite{FFMT} exposes the role
of the moduli fields and shows
that the SVD arises by an exchange of two Wilson lines. 
The SVD can be interpreted in terms of the breaking of
the $N=2$ worldsheet supersymmetry on the bosonic side of the
heterotic--string, and the map between
the dual vacua is induced in terms of a spectral flow
operator, which induces the map between the dual Wilson--lines. 
The bosonic representation of the 
Spinor--Vector Duality is useful 
for studying its imprint in
the Effective Field Theory (EFT) limit.

The freedrom in the choice of the discrete torsion can be translated
to two distinct choices of the Wilson--line background fields ref. \cite{FFMT}, 
which are mapped to each other by the spectral flow operator and
produce the spectra of the dual models.
The dual Wilson lines are given by 
\begin{align}\label{WL+}
		g = (0,0,0,0,0,1|0,0|1,0,0,0,0,0,0,0).
\end{align}
and 
\begin{align}\label{WL-}
	    g = (0,0,0,0,0,0|1,0|1,0,0,0,0,0,0,0),
\end{align}
whereas the
$Z_2$ twist of the internal coordinates is given by
eq. (\ref{z2twist}).
This picture again resembles that of $T$--duality in which the 
duality map corresponds to a map between two sets
of the metric and asymmetric tensor background fields.
The bosonic data in the form of eq. eqs. (\ref{WL+}) and (\ref{WL-})
can be used to study the imprint of the worldsheet symmetries
in the effective field theory limit (EFT) of the string compactifications, 
as well as the extension of the SVD to interacting CFTs \cite{AFG}.
This is obtained by using Gepner's construction \cite{gepner}
of (2,2) minimal models. Subsequently, breaking $(2,2)\rightarrow(2,0)$
and inducing the map with the spectral flow operator, revealing a map
between $(2,0)$ models that extends the SVD map. 

\section{SVD in the EFT limit}

The most important consequence of mirror symmetry is the implication
in the effective field theory limit of the string compactifications.
In the EFT limit, string vacua correspond to supergravity field theories
that encode the dynamics of the massless string modes. In this limit,
the ten dimensional supergravity theories are compactified on three
dimensional complex manifolds that preserve $N=1$ spacetime supersymmetry, 
{\it i.e.} on Calabi--Yau (CY) manifolds \cite{chsw}.
The worldsheet symmetry that corresponds to mirror symmetry implies
a similar symmetry in the EFT limit, {\it i.e.} it implies the existence
of Calabi--Yau manifolds that are related by the mirror symmetry map.
This in itself is already an unexpected result from the mathematical
point of view. However, the mirror symmetry map has additional
important consequences in the mathematical field of enumerative
geometry. It facilitates the counting of the intersection of the
curves in a given CY manfilold that are simplified on the dual
manifold. Thus, the innocuous worldsheet symmetry has profound
implications in the EFT limit of pure mathematical interest
in complex manifolds and their properties. 

Spinor--Vector Duality is a symmetry
in the space of (2,0) heterotic--string compactifications. It is a fascinating
symmetry because it exchanges representations that from the low
energy physical point of view play an entirely different role. 
The spinorial 16 of $SO(10)$ encode the Standard Model fermions,
whereas the vectorial 10 representation gives rise to the 
Standard Model Higgs multiplets. Yet, from the point of view
of the worldsheet string theory they are in a sense completely
equivalent. Furthermore, it is noted that $16\ne10$. Yet, the 10
always comes with additional 6 $SO(10)$ singlets that compensates
for this mismatch. What we learn is that the string theory does not 
care how the states rearrange themselves under the low energy EFT.
The string theory imposes that the correct number of states
is obtained in order to produce a modular invariant partition function. 

The interest here is in the relation to mirror
symmetry and the implication for the mathematical structures that
arise in the EFT limit. From the worldsheet string theory point of
view, SVD is a mundane extension of mirror symmetry. It corresponds 
to a smmetry under the mappings of the Wilson line moduli, rather
than the moduli of the internal compactified space. 
From the point of view of the EFT limit, it implies the existence 
of relations between supergravity compactifications on six dimensional
complex manifolds with vector bundles on them that correspond to the
gauge degrees of freedom of the heterotic--string. Furthermore, 
it can served as a tool in the analysis of complex manifolds with the
vector bundles. Furthermore, the existence of the Spinor--Vector Duality
in the ultra violet complete string theory raises the question
whether any effective field theory compactification of supergravity
on six dimensional manifold possess this property.
The Spinor--Vector Duality can therefore serve as a tool
to constrain the viable effective field theory limits of
string compactifications. In this respect, the relation with 
$T$--duality serves as a useful guide. In the case of $T$--duality
on a circle, we vary the moduli, which in this case is the radius
of the circle, pass through the self--dual point which is the 
enhanced symmetry point and emerge on the dual side.
In this case, the interpolation is continuous. If we project
the moduli with an orbifold the interpolation can no--longer be continuous. 
There is still, however, a discrete remnant of the symmetry under the 
exchange of the radius with its inverse. A similar phenomena occurs
with Spinor--Vector Duality. In the of vacua that preserve the
$N=2$ spacetime supersymmetry, the interpolation is continuous
as the Wilson--line moduli are retained in the spectrum, and 
at the self--dual point the symmetry is enhanced to $E_7$. 
In the vacua in which the $N=2~\rightarrow~N=1$ spacetime
supersymmetry by a second $Z_2$ orbifold twist, the Wilson--line
moduli are projected out and the map between the dual vacua
is discrete \cite{FFMT}. 

Spinor--Vector Duality may similarly have profound purely 
mathematical implications. It relates six dimensional compactifications
and the vector bundles on them via the Spinor--Vector Duality map. In refs. 
\cite{FGNHH1,FGNHH2, FGNHH3, FGNHH4} we initiated the exploration of this
relation in the EFT limit of the string compactifications. Our strategy
was to start with an orbifold that exhibits the SVD and resolve
the singularities using well established techniques in this context \cite{GNTW, GNHT},
producing a smooth compactification in the EFT limit. 
The SVD in the EFT limit of the string compactifications
in six and five dimensions was analysed in 
refs. \cite{FGNHH1} and \cite{FGNHH2}, respectively. 
SVD on $T^4/Z_2\times S^1$ in five dimensions
was analysed in ref. \cite{FGNHH1}, 
by including 
a Wilson line on $S^1$
in the form of eq. (\ref{WL+}) or (\ref{WL-})
and 
a twist in the form of eq. (\ref{z2twist})
that acts on four internal coordinates.
The next step is to analyse the
resolution of this orbifold to a smooth $K3\times S^1$ by
using some massless states in the orbifold model to resolve
the singularities. 
A discrete torsion incorporated in
the analysis of the orbifold model between the Wilson--line and
the twist and its effect on the resulting massless
states. 
The states used for the
resolution in the model that we analysed transform under the 
$SO(10)$ symmetry. 
This means
that the GUT group is broken by the resolutions. 
As the states available
for the resolution transform under the observable gauge symmetry
and differ in the dual configurations,
the gauge degrees of freedom are also different in the two dual
cases. 
The SVD in this case is therefore somewhat obscured. It is noted that 
some free fermionic models in which the $E_6$ symmetry is broken
to $SO(10)\times U(1)^3\times SO(16)$ give rise to states
in the vectorial 16 representation of the hidden $SO(16)$.
In these cases the hidden sector states can be used to
resolve the singularities without affecting the gauge group.
As the role of the discrete torsion in the smooth
EFT limit is obscured, an educated guess is made on
the resolved manifold of the orbifold with the discrete
torsion. 
The case without torsion is well defined in the resolved limit.
However, the case with torsion introduces some
subtleties that are discussed in ref. \cite{FGNHH1}. 
In short, the smooth geometries do exhibit a
spinor--vector duality--like phenomenon, but due to the
different states available for the resolutions on the
dual models, the gauge symmetries differ in the resolved
manifolds. 
This is expected to be common
in the resolved limit because of the different states
available for the resolution, {\it e.g.} vectorials in
one case and spinorials in the other, 
in the orbifold example discussed in \cite{aft, FFMT}.
In ref. \cite{FGNHH2} it was shown that the
spinor--vector duality operates 
in six dimensions as well.
In this case the vacua are self--dual under
the spinor--vector dulaity transformation, 
and satisfy a general anomaly
consistency condition on the number of 
spinor and vector representations
of any $SO(2N)$ unbroken subgroup in the string vacuum
\beq
N_V=2^{N-5}N_S+2N-8. \label{anomalycond}
\eeq
The analysis of SVD in the smooth $Z_2\times Z_2$
orbifold limit in four dimensions is more complicated
due to the large number of possible resolutions. 
The $T^6/Z_2\times Z_2$ orbifold has 64 $C^3/Z_2\times Z_2$ singularities, 
where $Z_2$--fixed tori intersect.
To produce a smooth manifold,
all the singularities have to be resolved, and
each singularity can be resolved in four
topologically distinct ways \cite{FGNHH3}. 
This gives rise to $4^{64}$ a priori disctinct possibilities. 
We can use the symmetry structure
of the $Z_2\times Z_2$ orbifold to reduce this number,
which still leaves $\sim10^{33}$ of distinct configurations. 
Many of the properties of the resolved EFT geometries, 
like the massless states and the interactions between them, 
depend on the chosen resolutions, and hinders the extraction of the
general properties of the resolved $Z_2\times Z_2$ orbifolds. 
A formalism that allows computations that are independent
of any choice of resolution was developed in
\cite{FGNHH3}. The analysis of SVD on resolved four dimensional
orbifolds is still outstanding.

Gauged Linear Sigma Models (GLSM) \cite{GLSM} provide another tool
in the analysis of the EFT limit of worldsheet string models.
The can be used to interpolate between the singular orbifold
models and their resolved smooth geometries.
Some properties of the worldsheet string vacua that
do not have a direct analogue in the smooth geometries can
be studied by using the GLSMs. A concrete example is
the discrete torsion in the worldsheet string partition functions
between the different modular orbits and has no direct analogue in the
smooth geometries that underlie the EFT limit. 
The GLSM was used in \cite{FGNHH4}
to reveal what becomes of the discrete torsion in the resolution
of the compact
$T^3/Z_2\times Z_2$ and non--compact $C^3/Z_2\times Z_2$ orbifolds. 
The GLSMs associated with the non–compact
orbifold with or without torsion are equivalent to a large degree:
only when expressed in the same superfield basis, a field
redefinition anomaly arises between them. In the orbifold limit
it reproduces the discrete torsion phases.
The GLSMs that correspond to the compact orbifold has 
mixed gauge anomalies, which need to
be cancelled by appropriate logarithmic superfield dependent
Fayet Iliopoulos–terms on the worldsheet, signalling
$H$–-flux due to NS5--branes supported at the exceptional cycles.

\section{From spectra to interactions} 

So far the discussion of mirror symmetry and spinor--vector duality was in terms of the 
relations between the massless states in the respective dual vacua, which are relations
at the level of one--loop worldsheet string amplitudes. The relations extend to
correlators between massless states in the string spectrum. The Yukawa couplings
are given in terms of correlators among
vertex operators
$$
\langle{V_1^fV_2^fV_3^b\cdot\cdot\cdot\cdot V_N^b\rangle},$$
where the vertex operators are given by \cite{calculating}
\beqn
V^{f}_{(-{1\over2})}& = &{\rm e}^{(-{c\over2})}~{\cal L}^\ell~%
{\rm e}^{(i\alpha\chi_{_{12}})}~
{\rm e}^{(i\beta \chi_{_{34}})}~
{\rm e}^{(i\gamma\chi_{_{56}})}~\nonumber\\
%
& & \left(~{\prod_{j}}{\rm e}^{(iq_i\zeta_{j})}~
~\{\sigma's\}~
{\prod_{j}}{\rm e}^{(i{\bar q}_i{\bar\zeta}_{j})}~\right)\nonumber\\
%
& & {\rm e}^{(i{\bar\alpha}{\bar\eta}_{1})}~
{\rm e}^{(i{\bar\beta}{\bar\eta}_{2})}~
{\rm e}^{(i{\bar\gamma}{\bar\eta}_{3})}~
{\rm e}^{(iW_R\cdot{\bar J})}
{\rm e}^{(i{1\over2}KX)}~
{\rm e}^{({\rm i}{1\over2}K\cdot{\bar X})}. \label{yukawas} 
\eeqn
The different components entering eq. (\ref{yukawas})
are discussed in \cite{calculating}.
The non--vanishing
correlators are invariant under all the string 
symmetries and selection rules.
In the vacua with enhanced $E_6$ symmetry, the couplings are
between three $27$ chiral representations. 
The mirror map implies that
$$~~~27\cdot27\cdot27~\longleftrightarrow~
\overline{27}\cdot\overline{27}\cdot\overline{27}$$
In the EFT limit the Yukawa couplings correspond to
intersection of curves on the Calabi--Yau manifolds. 
The worldsheet correlators have imprints in the geometrical data
and can be used to analyse the properties of the corresponding manifolds.
Much of the interest in mirror symmetry stems from its utility in the
purely mathematical field of enumerative geometry.
A very lucid introduction to the subject and its
connection to string theory is found 
in Sheldon Katz book \cite{Katz}. A more in depth treatment is found  in
\cite{mirrorsymmetry}. 
Briefly, the calculation of the intersection
of rational curves on CY manifolds is related to the calculation of
the Yukawa couplings. The Gromov--Witten invariants are related to 
the calculation of the Yukawa couplings and
provide a tool to analyse the geometrical data of the manifolds. 

To date, the analysis of Yukawa couplings in SVD dual string models
has not been performed even at a conceptual level. From
the decomposition of the chiral 27 respresentation of $E_6$
shown in eq. (\ref{e627tso10}) we note that the
couplings in terms of the $SO(10)\times U(1)$ $E_6$ subgroup are:
\begin{equation}
16_{+1/2}\cdot 16_{+1/2} \cdot 10_{-1} ~~~~
{\rm and}~~~ 10_{-1}\cdot 10_{-1} \cdot 1_{+2}
\label{svd27yuks}
\end{equation}
and 
\begin{equation}
\overline{16}_{-1/2}\cdot \overline{16}_{-1/2} \cdot 10_{+1} ~~~~
{\rm and}~~~ 10_{+1}\cdot 10_{+1} \cdot 1_{-2}
\label{svd27baryuks}
\end{equation}
for the $\overline{27}$. 
The relation of SVD to mirror symmetry, 
{\it i.e.} both represents mappings under
transformations of the parameters of
the Narain moduli spaces, suggests that SVD may
have similar interesting mathematical implications 
in the EFT limit. 
The likely tool is the calculation of Yukawa couplings
among the massless states as in mirror symmetry, albeit in the case of SVD
the picture is complicated because it involves the internal
manifold and the vector bundles on them. 

\section{Discussion and conclusion}

String phenomenology aims to connect between string theory and
observational data. These may come in the form of particle physics
data, astrophysical data, cosmological data and the emerging
field of gravitational waves. Ultimately, many of the signatures of string vacua, 
like the possible existence of extra vector bosons and supersymmetric
states, will be confronted with the observational data by using the tools
of point particle quantum field theories. It is therefore of vital importance
to know how to relate string vacua to their effective field theory limit.
Understanding the space of string compactifications and the symmetry
relations between them is another key element in the program to relate
string theory to observational data. The moduli of these spaces
underlie many of the symmetry relations in the space of string vacua, 
{\it e.g.} mirror symmetry arises from the exchange of the internal
moduli, the complex and K\"ahler strutcure moduli, whereas SVD arises
from the exchange of Wilson--line moduli. The expectation values
of the moduli fields also determine many of the parameters
in the low energy EFT. Understanding the moduli
spaces of the string compactifications is therefore key to understanding
their implications for both observational phenomena, as well as their
mathematical properties.

Spinor--Vector Duality arises from the breaking of $(2,2)$ worldsheet supersymmetry
to $(2,0)$, which if of interest from the phenomenological point of view . 
SVD provides a tool to study the moduli space of the $(2,0)$ heterotic--string
compactifications in the EFT limit. Similary to $T$--duality, the
point in the moduli space with $(2,2)$ worldsheet supersymmetry
is an enhanced symmetry point in the moduli space, where in the case
of SVD the moduli is a Wilson--line moduli. In the case of single
of a single $Z_2$--twist the deformation from the enhanced symmetry
point is continuous, whereas in the case of $Z_2\times Z_2$ the
Wilson--line moduli is projected out and the
SVD is induced by a discrete transformation.
In both cases the SVD dual string vacua
are connected to the enhanced symmetry point
with continuous or discrete transformations.
The question then is whether the SVD is complete, {\it i.e.}
does every viable EFT of quantum gravity have to be related
on a point in the moduli space with enhanced $(2,2)$ worldsheet
supersymmetry. 
We can pose a conjecture \cite{universe}: ``Every EFT
(2,0) heterotic--string compactification which has an ultra--violet
complete embedding in string theory is connected to a (2,2)
heterotic--string compactification by continuous interpolation or 
by orbifold''. Otherwise, it is necessarily in the ``Swampland''. 
The motivation stems from the question whether the symmetries of the string 
worldsheet theories are complete. We can regard it in analogy with the
$T$--duality and mirror symmetry, where similarly, we may
question whether $T$--duality is a complete symmetry of string theory, 
and whether a mirror manifold should always exist.
$T$--duality is a symmetry most readily associated with toroidal
compactifications. We can envision that the general theory of
quantum gravity must admit a symmetry that can be related
as $T$--duality and can be connected to a self--dual point.

The relation between quantum gravity EFTs and their realisation in ultraviolet
complete string theories constitutes much of the discourse in contemporary string 
phenomenology. A major focus of the research is on the question, which EFTs can
be realised in string theory versus those that do not.
The so--called "Swampland program" is a bottom--up approach to the
investigation of the relations between the quantum gravity EFTs and their
possible embedding in the ultra--violet complete theories of quantum
gravity, {\it i.e.} in string theory. Ultimately, the distinction between the
EFTs and the ultraviolet complete string theories is that in the EFTS only
the massless degrees of freedom are considered, whereas in the ultraviolet complete
string theories include the massive spectrum as well. In fact, the duality
transformations involve exchange of massless and massive modes in the
string spectrum. What looks mysterious from the EFT point of view
is apparent from the string theory perspective.
Mirror symmetry and SVD exemplify this point.
However, the symmetry structure underlying the worldsheet string theories
is far more complex and is related to the rich symmetry structure that
underlies their compactifications to two dimensions \cite{panoslattices, universe2022}.
Understanding the role of this rich symmetry structure
in the EFT limit has barely scratched the surface, 
let alone their possible phenomenological implications \cite{bsm2021}.
In this context, Spinor--Vector Duality is induced by a modular map, 
which refers to the operation of the spectral flow operator
on the bosonic side of the heterotic string. Similar modular
maps include the supersymmetry generator on the supersymmetric
side and the ${\tilde S}$ modular map in ref. \cite{stildemap, fmp}
between supersymmetric and non--supersymmetric string vacua.
Mirror symmetry and Spinor--Vector Duality offer a glimpse
into the relation between the ultra--violet complete string theories
and how their rich symmetry structure impacts the low energy effective
field theory limit. There is yet a much richer symmetry structure
that is yet to be explored in what can be termed in contemporary
discourse as "a top--down approach to the Swampland--program".

%
%

\end{document}